\documentclass[a4paper,12pt]{article}
\begin{document}
\hsize=16.0 truecm
\vsize=23.5 truecm
\baselineskip=13 pt
\newcommand{\M}{$\mu_B$}
\newcommand{\CMP}{{ Comm.\ Math.\ Phys.\ }}
\newcommand{\NP}{{ Nucl.\ Phys.\ }}
\newcommand{\PL}{{ Phys.\ Lett.\ }}
\newcommand{\PR}{{ Phys.\ Rev.\ }}
\newcommand{\PRep}{{ Phys.\ Rep.\ }}
\newcommand{\PRL}{{ Phys.\ Rev.\ Lett.\ }}
\newcommand{\RMP}{{ Rev.\ Mod.\ Phys.\ }}
\newcommand{\ZP}{{ Z.\ Phys.\ }}
\thispagestyle{empty}
\mbox{ }\hfill{\normalsize February 1996}\\
\mbox{ }\hfill{\normalsize CERN-TH/95-298}\\
\mbox{ }\hfill{\normalsize AZPH-TH/95-26}\\
\mbox{ }\hfill{\normalsize BI-TP 95/35} \\
\mbox{ }\hfill{\normalsize UCT-TP 95/225} \\
\pagenumbering{arabic}
\begin{center}
{\Large \bf THERMAL HADRON PRODUCTION\\[.5cm]
 IN Si-Au COLLISIONS}\\[1.cm]
{\bf J. Cleymans$^1$, D. Elliott$^1$, H. Satz$^{2,3}$ and R.L. Thews$^4$
}\\[2.0em]
$^1$Department of Physics,
University of Cape Town, Rondebosch 7700, South Africa  \\
$^2$Theory Division, CERN, CH-1211 Geneva 23, Switzerland \\
$^3$Fakult\"at f\"ur Physik, Universit\"at Bielefeld, 
D-33501 Bielefeld, Germany\\
$^4$Department of Physics,
University of Arizona, Tucson, AZ85721, USA
%
\begin{abstract}
The most abundantly produced hadron species in $Si\!-\!Au$ collisions
at the BNL-AGS (nucleons, pions, kaons, antikaons and hyperons)
are shown
to be in accord with emission from a thermal resonance gas source of
temperature $T\simeq 110$ MeV and baryochemical potential $\mu_B \simeq
540$ MeV, corresponding to about 1/3 standard nuclear density.
Our analysis takes the isopin asymmetry of the initial state
fully into account.
\end{abstract}
\end{center}
\newpage
\section{Introduction}
\medskip
The basic aim of high energy nuclear collisions is the production of
strongly interacting matter. Following the primary collision,
multiple scattering and production processes are expected to provide
the rapid increase in entropy needed for local equilibration. Hence the
first question to be addressed by nuclear collision experiments is
whether they indeed produce {\sl matter}, i.e., large-scale strongly
interacting systems in local equilibrium.
QCD thermodynamics predicts that at sufficiently
high density, such matter will be a plasma of deconfined quarks and
gluons (QGP). Since this prediction presupposes equilibrium,
experimental evidence for thermalisation is a prerequisite to
establishing QGP production.
\\
It would be ideal to test thermalisation at each stage in the
evolution of the produced systems. However, unambiguous tests exist
so far only for the final hadronic stage. In particular, we can check if
the abundance of hadrons emitted in nuclear collisions is governed
by thermal composition laws \cite{CS,CRSS}. The existence
of such a ``chemical equilibrium" means that the presence of different
hadron species at freeze-out is specified by the freeze-out
temperature $T$ and the freeze-out baryochemical potential
\M~determining the baryon density; hence the two parameters $T$ and
\M~fix the production ratios of all emitted hadrons, in much the same
way the abundances of light nuclei are determined by the conditions of
the early universe at the end of the strong interaction
era.\footnote{It has been noted
\cite{Heinz} that hadrons of a given species could be in local
{\sl thermal} equilibrium even though {\sl chemical} equilibrium
between the different species is not attained. }
If the initial state contains more neutrons than protons, it
is necessary to assure overall charge conservation, which leads,
e.g., to a $\pi^-/\pi^+$ ratio different from
unity.
The analy-\break sis presented below
takes such deviations from an isospin-symmetric initial state
fully into account. Since the neutron to proton
and the $\pi^-/\pi^+$ ratios deviate from one by about
ten to twenty percent, the results from this analysis will generally
differ from  the isospin-symmetric case
by this order of magnitude.
Estimates of the effects of a charge chemical potential have been
presented previously \cite{Gorenstein,Rafelski}.
\\
The confirmation
of chemical equilibrium among the emitted hadrons at freeze-out of
course does not imply that the previous stages were hadronic.
It is also not connected to any specific expansion
pattern, and in particular, it does not imply isotropic emission from a
single stationary ``fireball". Starting from a uniform medium,
hydrodynamic flow preserves chemical equilibration, but for anisotropic
boundary conditions, it leads to anisotropic momentum distributions.
Hence the momentum spectra of thermally emitted hadrons and the
source radii obtained from hadron-hadron interferometry
are needed to determine the nature of the expansion.
Hadron production according to thermal composition laws is thus
perfectly compatible with the existence of a QGP at an earlier point
in the evolution. It in fact provides a suitable basis for such a
possibility: if the final hadrons were emitted from an equilibrated
source, it is quite possible that this source was locally thermalised
already in its earlier stages and thus could have consisted of
deconfined matter.
\\
The experimental study of thermalisation is therefore of fundamental
importance. A recent comprehensive analysis of BNL-AGS data from
$Si\!-\!Au$
collisions \cite{BM-S,BM-S2} provides first
evidence that the most abundantly observed hadrons were indeed emitted
from a thermal source. If supported by further results from $Au\!-\!Au$
collisions at the AGS and at higher energies from the CERN-SPS, this
conclusion would be the
decisive step in showing that strongly interacting {\sl matter} can be
produced by nuclear collisions. In view of the importance of such a
result, we find it useful to analyse the present AGS data
independently in a self-contained hadrosynthesis approach. Our analysis
will therefore
be based solely on particle ratios, without any source volume or flow
features, and it will, as mentioned, take into account the initial state
isospin asymmetry.
\bigskip\noindent
\section{Hadronic Ratios.}
\medskip
The measured hadron production ratios listed in Table 1 form the basis
of our analysis. If the hadrons are emitted from a source in full
chemical equilibrium, each measured ratio determines a range of
$T$--\M~values with which it is compatible. This range can be calculated
on the basis of an ideal gas of all observed hadrons and hadronic
resonances, requiring overall charge and strangeness conservation and
correct resonance decays \cite{CS}. Let us briefly sketch the method.
If at freeze-out we have an ideal gas of hadrons and hadronic
resonances, then the system is described by the partition function
\begin{equation}
\ln Z(T,\mu_B,\mu_S,\mu_Q) =
\sum_i
\lambda_B^{B_i}\lambda_S^{S_i} \lambda_Q^{Q_i}  W_i . 
\end{equation}
Here $W_i$ is the phase space factor for hadrons of species
$i$ (mesons, baryons and their antiparticles), with $S_i$, $B_i$ and $Q_i$
 denoting the
strangeness, baryon number and charge of the hadron in question. The
baryonic fugacity is defined as $\lambda_B=\exp (\mu_B/T)$, that
for the strangeness as $\lambda_S={\rm exp}(\mu_S/T)$,
and that for charge as $\lambda_Q={\rm exp}(\mu_Q/T)$.
 The phase space
factors are given by
\begin{equation}
W_i = {d_i m_i^2 V T\over 2 \pi^2} K_2\left({m_i\over T}\right),
\end{equation}
with $d_i$ denoting the spin degeneracy, $m_i$ the mass of hadron species
$i$; $V$ is the volume of the system. We include in
the sums in eq.\ (1) all hadrons listed in the latest Particle Data
Compilation \cite{PDB}, excluding charm and bottom resonances;
decays are taken into account with their experimental
branching ratios, with an educated guess being made whenever the
information on decays is not complete. The resulting table contains
479 entries. The partition
function (1) then determines all thermal properties of the system in
terms of the four parameters $T, \mu_B, \mu_S$ and $\mu_Q$. The
chemical potential for the strangeness, $\mu_S$, can be fixed by
requiring the overall strangeness of the system to vanish. Similarly,
the chemical potential for the charge, $\mu_Q$, is fixed by requiring
the charge/baryon ratio of the final state to be equal to that of
the initial state. If there is a one-stage freeze-out of all
thermal hadrons, then all production ratios ($\pi/p,~
K/\pi,~K/{\bar K},~\rho/\pi,~\phi/\pi,~Y/p,...$)
are given in terms of the two remaining parameters $T$,
and $\mu_B$. For example, we can use the measured $\pi^+/p$ and
$K^+/\pi^+$ ratios to fix the values of $T$ and $\mu_B$;
all other ratios are then predicted,
if the different particle species are indeed present according to their
equilibrium weights.
\\
To fix the overall charge/baryon, we thus have to estimate the
number of interacting protons and neutrons in the initial state. For a
central $A\!-\!B$ collision ($A\ll B$), the number $N_{\rm part}$
of participant nucleons is the sum of the nucleons in $A$
and those in a tube of radius $R_A$ through nucleus $B$,
\begin{equation}
N_{\rm part} \simeq [A + (\pi R_A^2) n_0 2 R_B]
\simeq [A+ 1.5 A^{2/3}B^{1/3}]. 
\end{equation}
Here $n_0=0.17$ fm$^{-3}$ is standard nuclear density; $R_A=1.12
A^{1/3}$, and similarly for $B$.
For $Si\!-\!Au$ collisions, we thus obtain $N_{\rm part} \simeq 108$.
With $Z/A=0.5$ for $Si$ and $Z/A=0.4$ for $Au$, this is made up of
$N_p \simeq 46$ protons and $N_n\simeq 62$ neutrons. We thus
have to fix the overall charge/baryon of the final state at 46/108 by
suitably adjusting the charge potential $\mu_Q$ at the temperature $T$
and baryochemical potential \M~ obtained from particle ratios.
As noted, the strangeness potential $\mu_S$ is fixed to give
the final state a vanishing overall strangeness.
\\
We begin with the most abundantly observed hadron species,
$p,~\pi^{\pm},~K^{\pm}$ and $\Lambda$, since these are most likely to
be thermalised. What we denote by $\Lambda$ will always include the
$\Sigma^0$, since the two are experimentally not separable.
>From the production rates of these six hadron species,
we get five independent ratios. As seen in Fig.\ 1, four of these
in fact cross in a
common region in the plane of temperature and baryochemical potential,
so that pions, nucleons, kaons, antikaons and lambdas are observed
according to their thermal weights. The freeze-out parameters thus
obtained are
\begin{equation}
T_F=110 \pm 5~{\rm MeV}~~~~~~~~~~\mu_B^F=540 \pm 20~{\rm MeV}.
\end{equation}
The fifth  ratio, which can be taken either
as the $K^-/\pi^-$ or the  $\pi^-/\pi^+$ ratio, is also in agreement
with the above values.

In Fig.\ 2 we show that when charge conservation is not 
enforced (i.e. the isospin symmetric case, with
$\mu_Q = 0$), there is never a consistency between the $\pi^+/p$ and
$K^+/K^-$ experimental ratios. The additional constraint of charge 
conservation, normalized by the initial state participating nucleons,
provides an overlap region for these experimental ratios, allowing a 
consistent set of thermodynamic parameters to be determined.

We have also checked that the finite volume corrections discussed
in \cite{BM-S} do not modify our results.
In Table 1, we list all thermal ratios as determined by these values,
together with the measured ratios. It is evident that the emission rates
of nucleons, pions,
kaons, antikaons and hyperons are correctly described by thermal
composition. The corresponding baryon density is
$n_B=0.055 \pm 0.025$ fm$^{-3}$ and
hence about 1/3 standard nuclear density, indicating considerable
expansion before freeze-out.
The freeze-out values (4) agree with those determined in
\cite{BM-S}; our temperature is slightly lower. In \cite{BM-S},
the temperature range for freeze-out was
fixed by a study of $\Delta$ production. However, the determination of
the $\Delta$ yield is quite delicate, based on pion transverse
momentum distributions at small $p_T$ and/or reconstruction over a
partially known background \cite{Hemmick}. To avoid the uncertainties
this leads to, we use the directly measured hadron ratios. For
the freeze-out parameters (3), we get an overall thermal
ratio $\Delta/N = 0.25\pm 0.02$.
\\
Given the initial state baryon number of 108, we can extract from the
ratios of Table 1 the thermal abundances of the different species;
they are listed in Table 2. Also shown there are
the experimental abundances, obtained in the same way from
the experimental ratios.
\\
Next we turn to less copiously produced hadron species. In Table 1, we
also list their measured ratios together with the corresponding
predictions from thermal emission at the freeze-out parameters (4).
We see that the value for $\phi$ production \cite{Wang} agrees well
0.1 $\phi$'s per event. Multiply strange baryons ($\Xi$) seem
experimentally somewhat more abundant than thermally predicted.
A definite disagreement with the thermal predictions is found
for antibaryons (${\bar p}$ and ${\bar \Lambda}$): these are
produced much more copiously than their thermal predictions.
Before discussing the possible origin of this, we consider the
onset of equilibrium for the different species.
\bigskip 
\noindent
\section{Thermalisation.}
\medskip
Thermalisation in the AGS energy range can lead to an increase or to a
decrease of hadron abundances relative to those measured in $p-p$ or
$p-A$ interactions. Let us note three examples in some detail.
\\
A well documented case is the
$K^+/\pi^+$ ratio: it grows from a $p-p$ value near 0.05 to a
four times larger thermal value, above 0.2, in central $Si-Au$ and
$Au-Au$ collisions \cite{Zajc,Grazyna}). Similarly, the
$\phi/\pi^+$ ratio increases by more than a factor two from its $p-p$
value to the thermal result found in $Si-Au$ data \cite{phi/pi}.
\\
In contrast, the number of pions produced per participating nucleon
decreases at AGS energy
towards its thermal value. In Fig.\ 3, we show the $\pi/N_{\rm part}$
ratio in $p-A$
collisions at 14.6 GeV beam momentum \cite{Gaz}. To obtain these values,
we approximate the pion multiplicity by
$3\langle h_-\rangle$, where $\langle h_- \rangle$ is the average
number of negative hadrons; $\pi/N_{\rm part}$ is defined as the
average number of participants in a $p-A$ collision,
\begin{equation}
N_{\rm part} = 1 + n_0 \pi r_0^2(3/4)(2 R_A), 
\end{equation}
where $r_0\simeq 0.85$ fm is the proton radius.
Eq.\ (5) thus counts the number of nucleons contained in a
tube of radius $r_0$ through the nucleus; the factor (3/4) comes from
averaging over the impact parameter.
The ratio $\pi/N_{\rm part}$ is seen to be $A$-independent; it remains
constant at $1.5\pm0.2$ over the whole range from $p-N$ to $p-Au$.
To obtain the corresponding experimental value of $\pi/N_{\rm part}$
for $Si-Au$ collisions, we multiply the measured $\pi^+/p$ ratio by
$3 \times 1.14$ (where the factor 1.14 accounts for the observed
$\pi^-/\pi^+$ ratio \cite{Johanna}) and divide it by 2.35 in accord with
46 participating protons and 62 neutrons.
We thus find a significantly lower value, $1.15 \pm 0.12$, which is
in accord with the thermal value $133/108 \simeq 1.23$. This lower value
is also supported by data from $Si-Al$ collisions \cite{Gaz}.
The ratio $\pi/N_{\rm part}$ thus decreases from its $p-A$ value of
about 1.5 to a somewhat lower thermal value.
\\
Because of enhanced annihilation chances, the ratio of antibaryon
to baryon production is also expected to decrease in a dense
baryonic medium. In support of this, the ${\bar p}/p$ ratio appears to
decrease in going from $p-p$ to $A-B$ \cite{Shiva}.
\\
There is also no particular reason why all ratios should approach
thermalisation by the same mechanism or at the same rate.
Pion-nucleon and pion-pion interactions result in more
abundant kaon production than nucleon-nucleon interactions;
thus an environment with more such collisions will drive the
kaon to pion ratio up. In nucleus-nucleus collisions,
some of the available energy is needed to thermalise the participating
nucleons and therefore is removed from meson
production \cite{Gaz-priv}; hence at the AGS, with very limited energy,
$\pi/N$ decreases to its thermal limit. As mentioned,
antinucleons are more readily absorbed in nuclear matter, so that the
${\bar p}/p$ and ${\bar \Lambda}/\Lambda$ ratios in nuclear collisions
are expected to be smaller than in $p-p$ interactions.
And in general one would expect the approach to be
slower for less copiously produced particles. The onset of
thermalisation will therefore be quite specific for different species,
making a final uniform thermal distribution all the more striking.
\\
In view of the general evolution towards thermalisation, it seems
misleading to single out enhanced kaon/pion or hyperon/nucleon ratios as
``strangeness enhancement". There are fewer pions per nucleon in
nucleus-nucleus collisions than in $p-p$ or $p-A$ interactions: this
alone would drive the $K/\pi$ ratio up,
even for constant kaon production.
Moreover, there are strange hadron species (e.g., antikaons) whose
thermal production rate is smaller than that in hadron-hadron
collisions, and there are non-strange species (e.g., $\Delta$'s) with
enhanced thermal production.
\\
We return now to the discrepancy between experimental and thermal
antibaryon production. The thermal rates are determined by the
dependence on the baryochemical potential $\mu_B$:
increasing $\mu_B$ at fixed temperature decreases
the ratio ${\bar p}/p$ as $\lambda_B^{-2}={\rm exp}(-2\mu_B/T)$.
However, such a suppression presupposes that the antibaryons can really
experience the thermal medium, and that appears not clear.
The fate of antiprotons in nuclear matter
has been studied in detail in low energy
$p-A$ interactions \cite{Vaisen}.
Here one also expects enhanced annihilation,
particularly in the production of very slow ${\bar p}$'s, which spend a
long
time in the medium. In contrast, one finds that over a large momentum
range down to 0.5 GeV/c, the antiprotons apparently
do not interact with the medium. This has been interpreted in terms
of a very long formation time for baryon-antibaryon
pairs in nuclear collisions, so that antiprotons emerge as well-defined
particles only after leaving the nucleus \cite{Kopelio}. It has also been
considered as the effect of a specific screening of antiprotons in
nuclear matter \cite{Kahana}. In any case, until
the considerable transparency of nuclear matter for antiprotons
is understood, it is not clear what role they will play in the
medium produced in nucleus-nucleus collisions. It thus seems
safest to exclude them from thermal considerations of AGS data,
which are in the energy range studied in the mentioned $p-A$
collisions \cite{Vaisen}.
\bigskip \noindent
\section{Summary}
\medskip
We have here addressed the question of chemical
equilibrium at freeze-out in a self-contained fashion, including the
isospin asymmetry of the initial state. Our conclusion,
in full agreement with that of \cite{BM-S}, is that in $Si-Au$
collisions at the AGS all copiously produced hadron species are emitted
in accord with their thermal weights, as calculated for an ideal gas of
hadrons and hadron resonances with the freeze-out parameters given by
eq.\ (4). Some less frequently produced species also agree with this.
Our conclusion is thus supported by six or seven independent and
directly measured hadron production ratios, so that it seems quite
well-founded. In contrast, the production of the (relatively rare)
antibaryons is not suppressed as much as expected in a baryon-rich
environment.This agrees with results from ${\bar p}$ production in $p-A$
studies.
\\
The freeze-out temperature determined here can in principle be
counterchecked by the measured transverse momentum spectra, provided we
know the expansion pattern. In the absence of any collective flow
effects, we would have for light hadrons ($m_{\pi}\simeq 0$)
\begin{equation}
\langle |p_T| \rangle \simeq 2T \simeq 210~{\rm MeV}, 
\end{equation}
which is definitely too small. As noted in \cite{BM-S}, this could be
taken as an indication of predicted flow effects \cite{Heinz}. However,
to establish conclusively the presence of such effects requires a more
complete analysis, comparing in particular the change in transverse
momentum spectra in going from $p-A$ to $A-B$ collisions. Although of
great interest, that is beyond the scope of the present note.
\bigskip
Stimulating discussions with M. Gazdzicki, D. Kharzeev,
R. Stock and in particular
with J. Stachel are gratefully acknowledged.

\vfill 
\eject
\begin{center}
Table 1:
\medskip
Ratios of Hadron Species in Si-Au Collisions at the AGS
\medskip
(Thermal parameters: $T=110\pm 5$ MeV, $\mu_B=540\pm 20$
MeV)\\ [0.5 cm]
\begin{tabular}{|c|c|c|} \hline 
~~~~~~~~~ &~~~~~~~~~~~~~~~~~~~~~ &~~~~~~~~~~~~~~~~~~~\\
 Ratio    & Experimental  &   Thermal  \\
~~~~~~ &~~~~~~~~~~~~~~~~~~~~~ &~~~~~\\
\hline
~~  &~~                      &~                       \\
$\pi^+/p$   &~~~$0.80 \pm 0.08$ & ~~$0.87 \pm 0.15$ \\
$K^+/\pi^+$ &~~~$0.19 \pm 0.02$ & ~~$0.21 \pm 0.02$ \\
~~  &~~                      &~                       \\
\hline
~~  &~~                      &~                       \\
$K^+/K^-$   &~~~$4.40 \pm 0.40 $ & ~~$4.51 \pm 0.62$ \\
$\Lambda/p$   &~~~$0.20 \pm 0.04$ & ~~$0.16 \pm 0.02$ \\
$K^-/\pi^-$    &~~~$0.035 \pm .005$ & ~~$0.038 \pm 0.006$ \\
~~  &~~                      &~                       \\
\hline
~~  &~~                      &~                       \\
$\Xi^-/\Lambda$  &~~~$(1.2 \pm 0.2)\times 10^{-1}$ & ~~$(4.9 \pm
0.5)\times 10^{-2}$
\\
$\phi/\pi^+$   &~~~$(4.5 \pm 1.2)\times 10^{-3}$ &
~~$(4.6 \pm 1.3)\times 10^{-3}$
\\
${\bar p}/p$  &~~~$(4.5 \pm 0.4)\times 10^{-4}$ &
 ~~$(7.2 \pm 6.3)\times 10^{-5}$
\\
${\bar \Lambda}/\Lambda$   &~~~$(2.0\pm 0.8)\times 10^{-3} $
&~~$(3.4 \pm 3.0)\times 10^{-4}$
\\
~~  &~~                      &~                       \\
\hline
\end{tabular}
\end{center}
\bigskip
\begin{center}
Table 2:
\medskip
Abundances of Hadron Species in Si-Au Collisions at the
AGS
\medskip
(Thermal parameters: $T=110\pm 5$ MeV, $\mu_B=540\pm 20$ MeV)\\ [0.5cm]
\begin{tabular}{|c|c|c|} \hline 
~~~~~~~    &~~~~~~~~~~     &~~~~~~~~~~~   \\
 Species   & Experimental  & Thermal      \\
~~~~~~     &~~~~~~         &~~~~~~~       \\ \hline
~~         &~~             &~~            \\
 nucleons  &     94        &           94 \\
 pions     & 120  & 133 \\
 kaons     &  14  &  17 \\
 hyperons  &  14  &  12 \\
 antikaons &   3  &   4 \\
 $\Xi$'s   &   2  &   1 \\
 $\phi$'s     & $2\times 10^{-1}$  & $ 2\times 10^{-1}$ \\
 antinucleons & $4\times 10^{-2}$  & $ 6\times 10^{-3}$ \\
 antihyperons & $3\times 10^{-2}$  & $ 4\times 10^{-3}$ \\
~~  &~~              &~~                 \\  \hline
\end{tabular}
\end{center}
\vfill\eject
\centerline{Figure Captions:}
\bigskip
\begin{itemize}
\item{Fig. 1:}{The $T-\mu_B$ regions determined by the indicated
particle ratios (including experimental errors). The charge
is kept fixed at 46, corresponding to 46 participant protons
and 62 participant neutrons.}
\medskip
\item{Fig. 2:}{The $T-\mu_B$ regions determined by the indicated
particle ratios (including experimental errors). The charge
chemical potential is taken to be zero, thus neglecting the isospin
asymmetry of the initial state.}
\medskip
\item{Fig. 3:}{The ratio $\pi/N_{\rm part}$ as function of the number of
participating nucleons $N_{\rm part}$ in $p-A$ and in $Si-Al$ and
$Si-Au$ collisions at the AGS.}
\end{itemize}
\end{document}